\documentclass[aps,prl,twocolumn,superscriptaddress]{revtex4-1}

\usepackage{graphicx}
\usepackage{ulem}
\usepackage{color}

\usepackage{xspace}

\usepackage{amsmath}
\usepackage{siunitx}
\usepackage{miller}

\usepackage[version=3]{mhchem}
\usepackage{natbib}

\newcommand{\bk}[1]{{\langle #1 \rangle}}
\newcommand{\ten}[1]{{\boldsymbol #1}}
\renewcommand{\vec}[1]{{\boldsymbol #1}}

\begin{document}


\title{Interplay of electronic and spin degrees in ferromagnetic SrRuO$_3$: anomalous softening of magnon gap and stiffness}

\author{K. Jenni}
\affiliation{$I\hspace{-.1em}I$. Physikalisches Institut,
Universit\"at zu K\"oln, Z\"ulpicher Str. 77, D-50937 K\"oln,
Germany}

\author{S. Kunkem\"oller}
\affiliation{$I\hspace{-.1em}I$. Physikalisches Institut,
Universit\"at zu K\"oln, Z\"ulpicher Str. 77, D-50937 K\"oln,
Germany}

\author{D. Br\"uning}
\affiliation{$I\hspace{-.1em}I$. Physikalisches Institut,
Universit\"at zu K\"oln, Z\"ulpicher Str. 77, D-50937 K\"oln,
Germany}

\author{T. Lorenz}
\affiliation{$I\hspace{-.1em}I$. Physikalisches Institut,
Universit\"at zu K\"oln, Z\"ulpicher Str. 77, D-50937 K\"oln,
Germany}

\author{Y. Sidis}
\affiliation{Laboratoire L\'eon Brillouin, C.E.A./C.N.R.S., F-91191 Gif-sur-Yvette CEDEX, France}

\author{A. Schneidewind}
\affiliation{J\"ulich Centre for Neutron Science (JCNS) at Heinz Maier-Leibnitz Zentrum (MLZ), Forschungszentrum J\"ulich GmbH, Lichtenbergstra�e 1, 85748 Garching, Germany}

\author{A. A. Nugroho}
\affiliation{ Faculty of Mathematics and Natural Science, Institut Teknologi Bandung, Jalan Ganesha 10, 40132 Bandung, Indonesia}

\author{A. Rosch}
\affiliation{Institut f\"ur Theoretische Physik,
Universit\"at zu K\"oln, Z\"ulpicher Str. 77a, D-50937 K\"oln,
Germany}
\author{D.I. Khomskii}
\affiliation{$I\hspace{-.1em}I$. Physikalisches Institut,
Universit\"at zu K\"oln, Z\"ulpicher Str. 77, D-50937 K\"oln,
Germany}

\author{M. Braden}\email[e-mail: ]{braden@ph2.uni-koeln.de}
\affiliation{$I\hspace{-.1em}I$. Physikalisches Institut,
Universit\"at zu K\"oln, Z\"ulpicher Str. 77, D-50937 K\"oln,
Germany}





\date{\today}

\begin{abstract}

The magnon dispersion of ferromagnetic SrRuO$_3$ was studied by inelastic neutron scattering experiments on single crystals
as function of temperature.
Even at low temperature the magnon modes exhibit substantial broadening pointing to strong interaction with charge carriers.
We find an anomalous temperature dependence of both the magnon gap and the magnon stiffness, which soften upon cooling in the ferromagnetic phase.
Both effects trace the temperature dependence of the anomalous Hall effect.
We argue that these results show that Weyl points and the anomalous Hall effect can directly influence the spin dynamics in metallic
ferromagnets.
\end{abstract}

\pacs{}

\maketitle


Strong spin-orbit coupling (SOC) causes intertwining of charge and spin degrees of freedom, which may result
in various fascinating phenomena such as Weyl semi-metals \cite{Burkov2011,Wan2011,Murakami2007}, multiferroics \cite{Khomskii2009} or spin liquids with exotic excitations\cite{Kitaev2006,Chaloupka2010}.
For a ferromagnetic metal the combination of magnetic exchange with strong SOC splits the bands and causes the
emergence of Weyl points. These Weyl points possess strong impact not only on the charge transport \cite{Fang2003} but also on the
magnetic properties. The anomalous Hall effect can capture the impact of the Weyl points on the charge dynamics.
As we will show below, the same physics influences also directly the magnon dispersion, the magnon anisotropy-gap \cite{Itoh2016} and stiffness.

SrRuO$_3$ \cite{Randall1959,AlanCallaghan1966,Koster2012} with the 4d ion Ru$^{4+}$ is a prime candidate to observe the impact of strong SOC in a ferromagnetic metal.
It crystallizes in the cubic perovskite structure but undergoes two structural phase transitions at 975 and 800\,K into
an orthorhombic phase (space group Pnma) associated with rotations of the RuO$_6$ octahedra \cite{Randall1959,Jones1989,Chakoumakos1998}.
The ferromagnetic transition occurs at 165\,K in single crystals and at low temperature a magnetization of $\sim$1.6$\mu_B$ is observed
at 6\,T \cite{Kunkemoeller2016}. 
The electric
resistivity is linear at moderate temperatures and breaks the Ioffe-Regel limit, but it drops in the ferromagnetic phase with a clear kink at the ferromagnetic $T_c$ \cite{Allen1996,Klein1996} attaining very low residual resistivity values of only 3 $\mu \Omega cm$ in high-quality single crystals \cite{Kunkemoeller2016,Kunkemoeller2017b}.
This coupling of ferromagnetic excitations and charge transport inspired the proposal of spin-triplet pairing in the superconducting sister compound Sr$_2$RuO$_4$ \cite{Rice1995}, in which quasiferromagnetic  excitations could only recently be observed in neutron experiments \cite{Steffens2019}.
The anomalous Hall effect in SrRuO$_3$ shows a peculiar temperature dependence undergoing a sign change \cite{Izumi1997,Fang2003,Kats2004,Haham2011,Koster2012,Itoh2016}.
Fang et al. \cite{Fang2003} proposed that the anomalous Hall effect in SrRuO$_3$ arises from the impact of magnetic monopoles in momentum space associated with
Weyl points. The magnetic exchange splitting combined with the impact of SOC causes Weyl points in the band structure\cite{Fang2003,Itoh2016}, which DFT calculations
predict to occur near the Fermi level \cite{Chen2013}.

More recently Itoh et al. argued that the Weyl points not only induce the peculiar temperature dependence of the
anomalous Hall effect but also affect the spin dynamics. At each Weyl point, index $i$, two bands cross and the 
single-particle current $j^{W,i}_n=e \sum_{m} v^{i}_{nm}   \ten{\sigma}^m$ and magnetization $m^{W,i}_n=\sum_{m} g^{i}_{nm}  \mu_B \frac{\hbar \ten{\sigma}^m}{2}$
are proportional to each other \cite{Burkov2011} \begin{eqnarray}\label{magWeyl}
\vec{m}^{W,i}=\frac{\hbar \mu_B}{2 e} \ten{g}^i (\ten{v}^i)^{-1}\,  \vec{j}^{W,i}
\end{eqnarray}
where $\ten{g}^{i}$ is the $g$-tensor, $n,m=x,y,z$ and $\ten{v}^{i}$ the tensor of Fermi velocities characterizing the Weyl point $i$. The intimate relation of current, magnetization and Berry phases will be tested by our paper.

Inelastic neutron scattering (INS) experiments in reference \cite{Itoh2016} indeed find a temperature dependence of the magnon gap resembling that of
the anomalous Hall effect while the magnon stiffness was claimed to exhibit a normal temperature dependence, i.e. a hardening upon cooling.
However, these measurements were performed with powder samples that do not give direct access to the parameters of the magnon dispersion.
Here we report on INS, magnetization and anomalous Hall effect measurements on single-crystalline SrRuO$_3$. We confirm a close similarity
between the temperature dependencies of the magnon gap and the anomalous Hall conductivity, but the magnon gap differs from the powder experiment
by almost a factor two. Furthermore, the magnon stiffness also softens upon cooling in the ferromagnetic phase. We argue that this unusual temperature dependence originates from the coupling of current and magnetization
thus confirming the impact of the anomalous Hall effect on magnetization dynamics.

 \begin{figure}
 \includegraphics[width=\columnwidth]{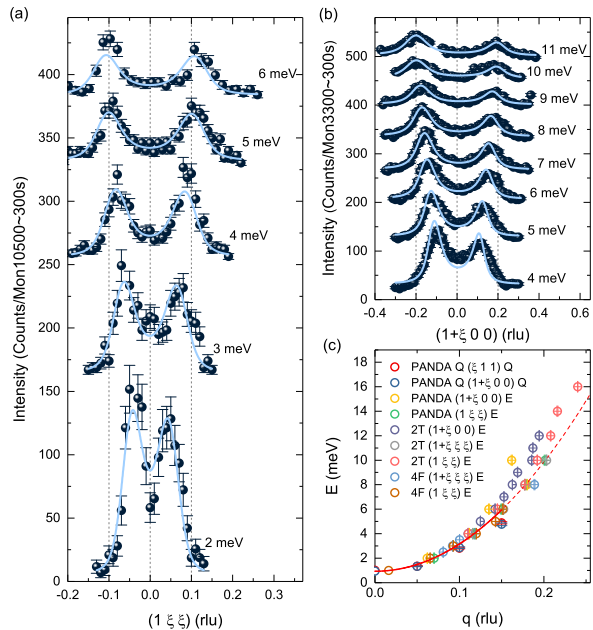}
  \caption{\label{structure}(a) and (b) Constant energy scans across the magnon dispersion in SrRuO$_3$ at various energies obtained at T=10\,K on the cold TAS 4F and on the thermal TAS 2T, respectively. Lines represent the folding of the modeled magnon dispersion including a broadening term with the resolution function of the spectrometer; data are vertically offset for clarity. (c) Combined
  magnon dispersion traced against the length of the magnon propagation vector $|q|$ in absolute units. The line reproduces the quadratic low $q$ behavior
  defining the magnon stiffness.}
  \label{disp}
 \end{figure}

Large single crystals of SrRuO$_3$ were grown by the floating-zone technique and characterized by resistivity and magnetization measurements \cite{Kunkemoeller2016}.
For the INS experiments six crystals with a total mass of $\sim$6g were coaligned in the [100]/[011] scattering plane in respect to the pseudocubic lattice with $a_c \sim 3.92$\,\AA . INS experiments were performed on cold triple-axis spectrometers (TAS) 4F at the Laboratoire L\'eon Brillouin and PANDA at the Meier-Leibnitz Zentrum, and on the thermal TAS 2T at the Laboratoire L\'eon Brillouin.
The anomalous Hall effect was measured on a rectangular sample with
edge lengths of $2.2 \times 1.64 \times 0.193$~mm$^3$ along the cubic
directions [1$\overline{1}$0], [001], and [110], respectively.
We applied the electrical current $I_x$ (typically 5~mA)  along
[1$\overline{1}$0] (orthorhombic $a$), the external magnetic field along
[110] (the easy axis, orthorhombic $c$, up to $\pm 7$~T) and measured the longitudinal ($U_x ||I_x$) and
the transverse voltage $U_y || [001]$ (orthorhombic $b$).  Using a SQUID
magnetometer, we also measured the magnetization $M$ of this sample for
the same field direction in order to precisely determine the normal and anomalous
Hall effects. Further experimental details are given in the supplemental material.\cite{suppl-mat}

 \begin{figure}
 \includegraphics[width=\columnwidth]{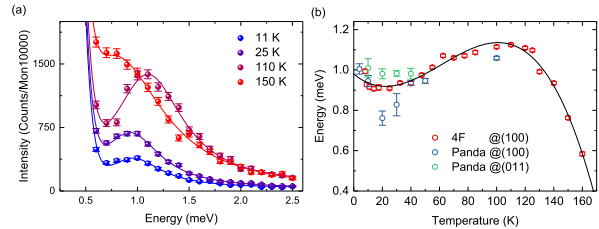}
  \caption{\label{structure}(a) Energy scans at the magnetic (and nuclear) Brillouin zone center (100) across the magnon gap arising from anisotropy. The data are taken
  at the cold TAS 4F with final neutron energy of 4.98\,meV. (b) The resulting gap values as function of temperature. Data taken on the cold TAS PANDA taken on an untwinned crystal are included.}
  \label{gap}
 \end{figure}

Fig. 1 shows the INS data obtained by constant energy scans across the magnon dispersion on cold and thermal TAS. The peaks arising from the magnon on both sides
of the (100) Bragg points are clearly visible and allow for a reliable determination of the dispersion.
In order to quantitatively analyze this data we calculate the folding of the magnon dispersion with the resolution
of the cold and thermal TAS using the ResLib program package \cite{Reslib}.
The lines superposed to the constant energy scans refer to this folding procedure with only a few global fit parameters.
For small momenta, the magnon dispersion can approximately be described by
$ E_0({\vec q})\approx\Delta + D \, {\vec q}^2$,
with the anisotropy gap $\Delta$ and the magnon stiffness $D$. 
Magnetic anisotropy is sizeable in SrRuO$_3$
as it can be inferred from the macroscopic magnetic anisotropy \cite{Kanbayasi1976,Cao1997,Kunkemoeller2016} and the shape memory effect \cite{Kunkemoeller2017b} and from an optical study \cite{Langner2009}.
The small orthorhombic distortion of SrRuO$_3$
induces a tiny anisotropy in $D$, see below.
Note, that throughout the paper we use reduced reciprocal lattice units with respect to the pseudocubic cell, $\frac{2\pi}{a_c}$,  but the stiffness is typically given in units of meV\AA$^2$. In order to describe the measured scan profiles we need to assume a width of the magnon modes that amounts to 40\% of their energy. Part from this broadening can stem from the twinning of the crystals superposing different direction of the orthorhombic lattice, but due to
the small orthorhombic distortion this broadening should be of the order of a few \% only.
Magnons in SrRuO$_3$ thus exhibit strong scattering most likely due to the coupling to electrons, see also the kink in resistivity at $T_c$ \cite{Allen1996,Klein1996}, and
due to the presence of sizable SOC.
The dispersion, which for small $|\vec q|$ values is quadratic, is presented in Fig. 1(c).
A consistent description of the data at low temperatures is obtained yielding $D$=87(2)\,meV\AA$^2$ and $\Delta$=0.94(3)\,meV and an energy width of $0.4\cdot E_0(q)$. Note that the gap $\Delta$ is almost a factor two smaller than the result obtained from the powder experiment, and also the stiffness $D$, considerably differs \cite{Itoh2016}.

\begin{figure}
\includegraphics[width=\columnwidth]{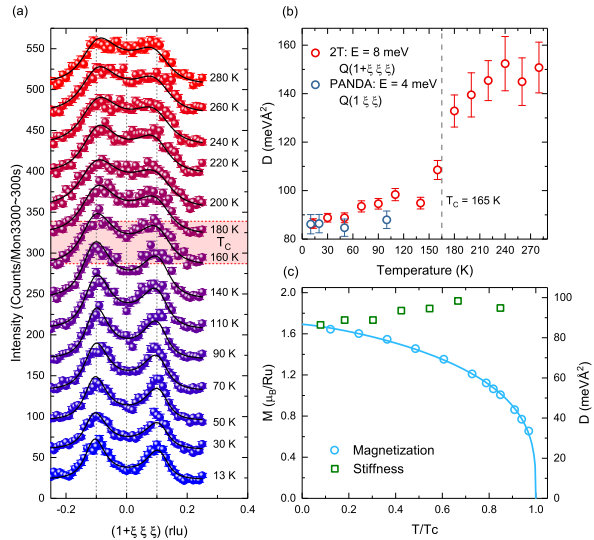}
  \caption{\label{structure}(a) Constant energy scans across the magnon dispersion in SrRuO$_3$ at 8\,meV obtained at the  thermal TAS 2T. Lines represent the folding of the magnon dispersion including a broadening term with the resolution function of the spectrometer. (b) Temperature dependence of the magnon stiffness obtained from
  the data in (a) and from additional measurements on the cold TAS PANDA. Above the ferromagnetic transition $D$ just describes the position of the paramagnon scattering, see text. (c) Comparison of the temperature dependencies of the magnetization and that of the magnon stiffnes.}
  \label{Tdep}
 \end{figure}

With the better resolution of the cold TAS 4F we scanned across the magnon gap at the zone center, see Fig. 2. Again the gap can be directly read from the raw data in contrast to the previous powder INS experiment.\cite{Itoh2016}
Our single-crystal INS result further agrees with the optical study \cite{Langner2009} and with the extrapolation of the anisotropic magnetization \cite{Kunkemoeller2017b}. The easy axis of SrRuO$_3$ is found along
the orthorhombic $c$ direction in Pnma notation \cite{Kunkemoeller2017b} and therefore two anisotropy gaps can be expected for the orthorhombic system, i.e. for rotating the magnetic moments towards $a$ and $b$ directions.
Our data indicate little splitting for the two gaps that were examined with an untwined smaller crystal on PANDA.
This crystal was first mechanically detwinned \cite{Kunkemoeller2016} and then mounted with its cubic [01$\bar{1}$] direction parallel to a magnetic field in a vertical field cryostat. After applying a magnetic field of 3\,T a fully monodomain crystal was obtained \cite{Kunkemoeller2017b} with the orthorhombic $c$ axis, the magnetic easy axis, parallel to cubic [01$\bar{1}$], and thus $a\|$[011] and $b\|$[100]. The gaps measured at the scattering vectors $\bf{Q}$=(100) and (011) correspond thus to the rotation of the moments
towards $a$ and $b$, respectively. In agreement with the macroscopic analysis in reference \cite{Kunkemoeller2017b} the $a$ direction is only slightly softer than $b$.
The anomalies of the gap temperature dependence are qualitatively confirmed by these monodomain measurements, but the temperature dependence of the averaged gap
profits from higher statistics.
From the constant $\mathbf{Q}$ scans at the zone center we
deduce the temperature dependence of the anisotropy gap by fitting a gaussian peak, see Fig. 2 (b).
In only qualitative agreement with the powder INS there is
a rather anomalous softening and rehardening of the gap upon cooling deep in the ferromagnetic phase, while the closing of the gap upon heating above the Curie
temperature corresponds to the expected behavior.

Fig. 3 summarizes the temperature dependence of the magnon stiffness.
We recorded constant energy scans at 8\,meV  between 13 and 280\,K  that were analyzed by fitting the $D$ value through the folding of the resolution
with the dispersion.
The characteristic two peak structure remains visible even well above the Curie temperature, which underlines
the persistence of ferromagnetic correlations. This scattering agrees with the expectation for a nearly ferromagnetic
metal, which still exhibits a paramagnon signal very similar to that of ferromagnetic material with broad magnons \cite{Moriya1985}.
The fitted $q$ positions can be directly transformed into temperature dependent stiffness constants, $D$, by taking
the temperature dependent anisotropy gaps into account,  Fig. 3 (b).
The magnon stiffness
clearly softens upon cooling well below the Curie temperature, while the magnetization increases.
In Fig. 3 (c) we compare the stiffness to the spontaneous magnetization,
because in a usual system one expects the two quantities to scale. In contrast in SrRuO$_3$ the stiffness softens upon cooling
well below T$_c$.
Scaling the magnetization, $M_s$, and $D$ by the low-temperature values, one recognizes that at 0.8 times $T_c$ the stiffness almost
twice as large than what follows from $D\propto m$.

 \begin{figure}
 \includegraphics[width=\columnwidth]{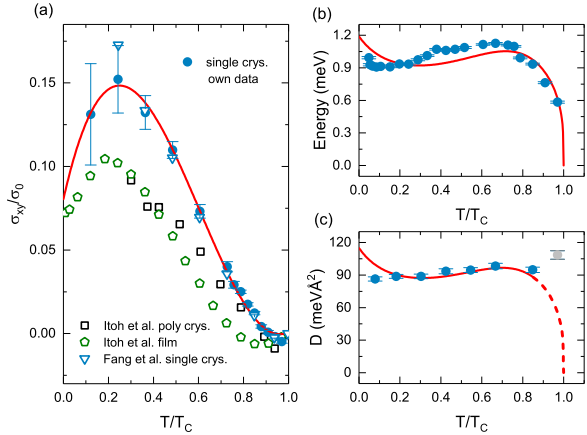}
  \caption{\label{structure} Comparison of the anomalous Hall effect with the magnon gap and stiffness. (a)  Temperature dependence of the anomalous Hall effect normalized
 to $\sigma_0$ measured on single crystals and analyzed with the magnetization curves obtained on the same samples \cite{noteAHE}.
 (b) Temperature dependence of the magnon gap (circles) compared to that of the anomalous Hall effect described by the relation (5).
 (c) Temperature dependence of the magnon stiffness analyzed in comparison to the anomalous Hall effect, relation (5).}
  \label{fit}
 \end{figure}

The  dynamics of small-amplitude, long-wavelength oscillations of the magnetization in a ferromagnet polarized in the $z$-direction
can be described by the action \cite{Itoh2016}
\begin{align}\label{Seff}
S&\approx \frac{1}{2} \int d^3 \vec r \,dt\, \alpha(T) (\frac{d m_x}{dt} m_y-\frac{d m_y}{dt} m_x)\\
& -\kappa(T) (m_x^2+m_y^2)-A(T) \!\left((\vec \nabla m_x)^2+(\vec \nabla m_y)^2\right)+\dots \nonumber
\end{align}
up to higher order corrections and damping terms. From the corresponding Euler-Lagrange equations one obtains
\begin{eqnarray}\label{deltaD}
\Delta=\frac{\kappa(T)}{\alpha(T)}, \qquad D=\frac{A(T)}{\alpha(T)}
\end{eqnarray}
In the absence of spin-orbit coupling $\alpha$ is exactly given by $\alpha_0=\frac{1}{2 m(T) \mu_B }$ where $m(T)$ is the magnetisation density. Taking SOC and the Weyl points into account $\alpha(T)$ is
determined not only by $m(T)$  but also by  the anomalous Hall effect \cite{Itoh2016,suppl-mat}
\begin{eqnarray}\label{alpha}
\alpha(T) \approx \frac{1}{g \mu_B m(T)} + c(T) \, \sigma^a_{xy}(T)
\end{eqnarray}
This equation arises because the time-dependent magnetization induces currents close to the Weyl points, see Eq.~\eqref{magWeyl} and \cite{suppl-mat}. Due to the Berry curvature of the Bloch bands, both a transverse current and a transverse magnetization are generated, modifying $\alpha(T)$.

Neglecting the $T$ dependence of  $\kappa(T)$, $c(T)$ and $A(T)$, we obtain from Eq.~\eqref{deltaD} and \eqref{alpha} for the spin gap $\Delta$ \cite{Itoh2016} and the
stiffness $D$ the same temperature dependence parametrized by
\begin{eqnarray}
\Delta(T)\approx \frac{a_\Delta m(T)/m_0}{1+b (m(T)/m_0)(\sigma_{xy}(T)/\sigma_0)} \nonumber \\
D(T)\approx \frac{a_D m(T)/m_0}{1+b (m(T)/m_0)(\sigma_{xy}(T)/\sigma_0)} \label{eq:1}
\end{eqnarray}
where $m_0=m(T=0)$ and $\sigma_0=\frac{e^2}{2 \pi \hbar a_c}$.

In order to verify the scaling between the anomalous Hall effect and the two characteristic parameters of the magnon
dispersion in SrRuO$_3$ we also measured the magnetization and anomalous Hall effect on a single crystal. The magnetization
obtained by extrapolating field-dependent magnetization curves to H=0\,T is shown in Fig. \ref{Tdep}(c) and can be described by a stretched power law $m(T)=m_0(1-(\frac{T}{T_c})^a)^\beta$ with
the parameters $m_0$=1.69$\mu_B$, $a$=1.27 and $\beta$=0.304. 
The field-dependent magnetization was inserted in the analysis
of the anomalous Hall effect measurement performed on the same sample.
The comparison of our and previous Hall conductivity results is shown in Fig. \ref{fit}(ab).
Due to the larger thickness of the single-crystalline sample the Hall voltage is smaller, but it offers the advantage
of more precise magnetization data.
Our single-crystal data of the Hall conductivity $\sigma_{xy}$ agree with previous single-crystal data
but only qualitatively with powder
and thin film data \cite{Itoh2016,noteAHE}. There is a sign change in $\sigma_{xy}$ slightly below the ferromagnetic transition.
Differences may stem from the twinning and different orientations of the distorted orthorhombic lattices in the powder and thin-film experiments and from differences in sample quality.
The temperature dependence of the anomalous Hall effect was fitted by a polynomial  and then
inserted in equation (\ref{eq:1}) to describe the anomalous softening of the magnon gap in the ferromagnetic phase, see Fig. \ref{fit}(b).
This analysis well reproduces the main feature with $a_\Delta$ = 1.66\,meV  and $b$ = 4.98.
The same temperature dependence of the anomalous Hall effect was used to describe the
softening of the stiffness compared to a normal behavior proportional to the magnetization, $a_D$ = 160.90\,meV\AA$^2$. Again good agreement is obtained, see Fig. \ref{fit}(c). Taking into account that Eq.~\eqref{eq:1} completely neglects corrections arising, e.g., from the $T$ and magnetization dependence of $K(T)$, $J(T)$, $c(T)$ and from the broadening of the spin-waves, the
semi-quantitative agreement is satisfactory. It clearly supports a common origin of the unusual softening of the spin-gap and the spin-stiffness towards lower temperatures, explained by the coupling of magnetization and current and by the $T$-dependence of the anomalous Hall effect.


At the phase transition, where $m(T)$ vanishes, Eq.~\eqref{eq:1} predicts that $D(T)$ vanishes also. However, in this temperature regime
the broad spin-waves smoothly transform into
paramagnon scattering  \cite{Moriya1985} with a very similar shape and thus a drop of $D(T)$ cannot be extracted from the measured neutron scattering data.

In conclusion we have studied the magnetization, anomalous Hall effect and magnon dispersion in SrRuO$_3$ using high-quality single crystals.
The magnon modes exhibit sizeable broadening revealing strong scattering by most likely charge carriers.
The magnon gap and the magnon stiffness do not follow the temperature dependence of the spontaneous magnetization. Both quantities soften upon cooling over a large temperature range in the ferromagnetic phase and
at least the magnon gap passes through a minimum. These findings can be well explained by the effect of Weyl points situated close to the Fermi level.
Such Weyl points possess a well-established impact on the anomalous Hall effect and cause an additional term in the magnetic Hamiltonian. The latter
leads to a reduction of both the magnon gap and the stiffness, which scales with the anomalous Hall effect. Our data perfectly agree with
this scaling between the anomalous Hall effect and the magnon dispersion.

\begin{acknowledgments}
We thank I. Lindfors-Vrejoiu for discussions.
This work was funded by the Deutsche Forschungsgemeinschaft (DFG,
German Research Foundation) - Project number 277146847 - CRC 1238, projects A02, B01, B04, C02, and C04.
\end{acknowledgments}

\bibliographystyle{apsrev4-1}

\begin{thebibliography}{29}
\bibitem [{not()}]{noteAHE}%
  \BibitemOpen
  \href@noop {} {}\bibinfo {note} {Note that there is an inconsistency in the
  sign of the AHE data published on SrRuO$_3$. In the low-temperature range the
  AHE is parallel to the normal Hall effect, which corresponds to electron
  charge carriers, and therefore $\sigma_{xy}\approx
  -\frac{\rho_{xy}}{\rho_{xx}^2}$ is positive in our notation.}\BibitemShut
  {Stop}%
\end{thebibliography}
%



\bigskip

\bigskip

\bigskip
\begin{center}
{\bf supplemental material}
\end{center}

- {\it INS experiments} - INS experiments were performed on cold triple-axis spectrometers (TAS) 4F at the Laboratoire L\'eon Brillouin and PANDA at the Meier-Leibnitz Zentrum, and on the thermal TAS 2T at the Laboratoire L\'eon Brillouin. All instruments were
equipped with pyrolithic graphite (002) monochromator and analyzer crystals. Scans were performed with constant final energies of 14.7 and 4.98\, meV and pyrolithic graphite and cooled Be filters were inserted behind the sample on 2T and 4F and PANDA, respectively.

- {\it AHE experiments} - The sample for the AHE measurements was cut from a larger single crystal that was mechanically
detwinned yielding typically $\approx 85\%$ single-domain
crystals \cite{Kunkemoeller2016,Kunkemoeller2017b}.
Fully detwinning is achieved when a magnetic field is applied
along the [110] direction, which then becomes the magnetic easy-axis
$c$.
In order to compensate the misalignment of the $U_y$-contact positions, we antisymmetrized the raw data of $U_y$
by calculating $U_H(B)=(U_y(+B)-U_y(-B))/2$ and derived the Hall resistivity $\rho_{xy}(B)=U_Hd/I_x$ where $d=0.193$~mm
denotes the sample thickness and $I_x$ the applied current of typically 5~mA. The normal and the anomalous Hall constants $R_N$ and $R_A$,
respectively, were determined from fits of $\rho_{xy}(B)= R_N B + R_A \mu_0M(B)$ using the $M(B)$ data obtained from magnetization
loops measured with a SQUID magnetometer on the sample sample for the same field orientation. The anomalous Hall resistivity
$\rho_{xy}(B=0) = R_A\mu_0M_s$ with the spontaneous magnetization $M_s$ is obtained through linear extrapolations $M_s=M(B\rightarrow 0)$ of the
measured $M(B)$ loops. In order to convert $\rho_{xy}(B=0)$ to the anomalous Hall conductivity $\sigma_{xy}= -\rho_{xy} /(\rho_{xx}^2+\rho_{xy}^2)$
we use the zero-field resistivity $\rho_{xx} = U_xA_x/(I_xd_x)$ where $d_x=0.64$~mm is the distance of the $U_x$ contacts and
$A_x=0.425$~mm$^2$ the sample's cross section.

- {\it Magnetization dynamics and calculation of  $\alpha$ } -
A description of the magnetization dynamics by a simple (classical) Heisenberg model $H=\sum_{ij}{J\mathbf{S}_i\cdot\mathbf{S}_j - K S_{i,z}^2}$ with anisotropy $K$ predicts $\Delta =K\bk{S_z}$ and $D=2J\bk{S_z}a_c^2$
with $J$=3.5(1)\,meV ($\bk{S_z}$=0.8 is estimated from the magnetization assuming a $g$ factor of 2). Such a description is, however, of limited validity for the metallic SrRuO$_3$ and cannot predict the temperature dependence of $\Delta$ and $D$. Instead, one can use the effective field theory, Eq.~\eqref{Seff} of the main text.

In the absence of spin-orbit coupling, the factor $\alpha(T)$ is exactly given by  $\alpha(T)=1/(2 \mu_B m(T))$. This  can be obtained from commutation relation $[S_x,S_y]= i \hbar S_z$, where the total spin $S_z$ can in the thermodynamic limit be replaced by $\langle S_z \rangle = V m(T)/(g \mu_B)$ with $g=2$ and the system volume $V$. In an itinerant system with spin-orbit coupling, $\alpha(T)$ arises from integrating out of the electrons and is given by the transverse qcomponent of the inverse of the bare electron susceptibility 
\begin{eqnarray}
\alpha = -\lim_{\omega \to 0,\vec q \to 0} \frac{(\chi_0^{-1}(\vec q,\omega))_{xy}}{i \omega}
\end{eqnarray}
$\alpha$ is therefore computed from a correlation function of the total magnetization of the conduction electron. Using that close to each Weyl nodes the electric current and the magnetization are proportional to each other, the total magnetization can be written as
\begin{eqnarray}
\vec m\approx\vec{m_0} + \frac{\hbar \mu_B}{2 e} \sum_i \ten{g}^i (\ten{v}^i)^{-1}\, (\vec j^{W,i} -\vec j^{W,\bar{i}})
\end{eqnarray}
where $\vec m_0$ are contributions to the magnetization not related to the Weyl points, while we used  Eq.~\eqref{magWeyl} (main text) to express the magnetization close to the Weyl points by current operators. Here we use the SrRuO$_3$ is inversion symmetric. Weyl points therefore occur in pairs connected by inversion symmetry which are denoted by  the indices $i$ and $\bar i$. Inversion symmetry maps $\vec j^{W,i}$ to $-\vec j^{W,\bar{i}}$.
Weyl points can be viewed as source of Berry curvature in momentum space, which is the origin of the anomalous Hall effect and thus are expected to give the main contribution to this quantity and its temperature dependence. The anomalous Hall conductivity is given the current-current correlation function
$\sigma^a_{xy}= \frac{\langle j_x j_y\rangle_\omega}{i \omega}\approx \sum_{i} \frac{\langle j^{W,i}_x j^{W,i}_y\rangle_\omega}{i \omega}$, where we assumed that it is dominated by contributions close to the Weyl points and that corrections from scattering between different Weyl nodes give only small contributions to the anomalous Hall effect. Under these conditions
\begin{eqnarray}\label{alpha}
\alpha(T) \approx \frac{1}{g \mu_B m(T)} + c(T) \, \sigma^a_{xy}(T)
\end{eqnarray}
where we approximate the contribution from $\vec m_0$ by its value without spin-orbit coupling. The constant
$c(T)\sim (\hbar \mu_B g/(e v \chi_{0,xx}))^2$ is proportional to  $\chi_{0,xx}^{-2}$. 
The bare transverse susceptibility is approximately given by $\chi_{0,xx} \approx m/B_{ex}$ where $B_{ex}\sim J m$ is the exchange field. Therefore $c(T)$ is expected to depend only weakly on temperature.

The physical mechanism behind Eq.~\eqref{alpha} is that the time-dependent magnetization induces forces on the electrons which can be described as electric fields proportional to $d\vec m/dt$. These forces have opposite sign on Weyl fermions at momenta related by inversion symmetry, thus the net current vanishes. The Berry curvature of the Bloch bands, however, has the effect that $d\vec m_x/dt$ induces at each Weyl node currents in the $\pm y$ direction inducing a net magnetization $m_y$. This mechanism leads to a correction to $\alpha$ and thus, according to Eq.~\eqref{deltaD}, to corrections to the spin stiffness and the spin gap.

\end{document}